\documentclass[aps,twocolumn]{revtex4}
\usepackage{graphicx}
\usepackage{latexsym}

\newcommand{\be}{\begin{equation}}
\newcommand{\ee}{\end{equation}}
\newcommand{\bea}{\begin{eqnarray}}
\newcommand{\eea}{\end{eqnarray}}

\begin{document}

\title{How do proteins search for their specific sites on coiled or globular DNA}
\author{Tao Hu, A.Yu.Grosberg, B.I.Shklovskii}
\affiliation{Department of Physics, University of Minnesota \\
116 Church Street SE, Minneapolis, MN 55455}
\date{\today}

\begin{abstract} It is known since the early days of molecular biology that
proteins locate their specific targets on DNA up to two orders of
magnitude faster than the Smoluchowski 3D diffusion rate.  It was
the idea due to Delbr\"uck that they are non-specifically adsorbed
on DNA, and sliding along DNA provides for the faster 1D search.
Surprisingly, the role of DNA conformation was never considered in
this context.  In this article, we explicitly address the relative
role of 3D diffusion and 1D sliding along coiled or globular DNA
and the possibility of correlated re-adsorbtion of desorbed
proteins. We have identified a wealth of new different scaling
regimes. We also found the maximal possible acceleration of the
reaction due to sliding, we found that the maximum on the
rate-versus-ionic strength curve is asymmetric, and that sliding
can lead not only to acceleration, but in some regimes to dramatic
deceleration of the reaction.  %We also discuss possible roles of molecular motors
%in the light of our findings.
\end{abstract}

\maketitle

\section{Introduction}

\subsection{The problem}

Imagine that while you are reading these lines a $\lambda$-phage
injects its DNA into a cell.  For the infected cell, this sets a
race against time: its hope to survive depends entirely on the
ability of the proper restriction enzyme to find and recognize the
specific site on viral DNA and then cut it, thus rendering viral
DNA inoperable and harmless. If restriction enzyme takes too long
to locate its target, then the cell is dead.

This is, of course, just an example.  Essentially all of molecular
biology is about various enzymes operating with the specific
places on DNA, and each enzyme must locate its target site quickly
and reliably.  How can they accomplish the task?  It was
recognized very early on that the search by free diffusion through
the 3D solution is far too slow and proteins somehow do it faster.
Indeed, the rate at which diffusing particles find the target was
determined by M. Smoluchowski as early as in 1917
\cite{Smoluchowski}, it is equal to $4 \pi D_3 b c$, where $b$ is
the target radius, $D_3$ and $c$ are, respectively, the diffusion
coefficient and concentration of diffusing particles, in our case
- proteins (see also appendix \ref{sec:Smoluchowski} for a simple
derivation).  Although Smoluchowski result sets the rigid upper
bound for the possible diffusion controlled rate, proteins at
least in some instances somehow manage to do it up to about two
orders of magnitude faster - see, for instance,
\cite{Riggs,Eigen}. The idea to resolve this paradox goes back to
Delbr\"uck \cite{Delbruck} who suggested that proteins can fairly
quickly adsorb on a non-specific random place on DNA and then 1D
sliding along DNA can be much faster than the 3D diffusion.  In
fact, the idea that reduced dimension speeds up chemical reaction
can be traced even further back to Langmuir \cite{Langmuir}, who
noticed that adsorbtion of reagents on a 2D surface can facilitate
their diffusive finding each other.

The field attracted intensive attention for many years.  Early
studies \cite{Riggs,Eigen} seemed to corroborate the Delbr\"uck
model. A nice recent review of various strategies employed to
address the problem experimentally can be found in the paper Ref.
\cite{Halford}.  Based on the summary of experimental evidence,
authors of this review conclude, that the process is not just the
naive 1D sliding, but rather a delicately weighted mixture of 1D
sliding over some distances and 3D diffusion.  A theorist also
could have guessed the presence of a cross-over between 1D sliding
and 3D diffusion, because sliding along coiled DNA becomes very
inefficient at large scale:  having moved by about $t^{1/2}$ along
DNA after 1D diffusion over some time $t$, protein moves in space
by only $t^{1/4}$ if DNA is a Gaussian coil. This is very slow
subdiffusion. That is the situation requiring theoretical
attention to understand how 3D and 1D diffusion can be combined
and how their combination should be manifested in experiments.

On the theoretical front, major contribution to the field is due
to Berg, Winter and von Hippel (BWH) \cite{BWH}. As an outcome of
their theory, these authors formulated the following nice
prediction, partially confirmed by their later \emph{in vitro}
experiments \cite{BWH2}: the rate at which proteins find their
specific target site on DNA depends in a non-monotonic fashion on
the ionic strength of the solution.  In this context, ionic
strength is believed to tune the strength of non-specific
adsorbtion of proteins on DNA, presumably because a protein
adsorbs to DNA via positively charged patch on its surface. Thus,
in essence one should speak of the non-monotonous dependence of
the rate on the energy of non-specific adsorbtion of proteins on
DNA.

Although qualitatively consistent with experiment, BWH theory
\cite{BWH} leaves several questions open.  First and foremost, how
does the search time of proteins finding their target, or the
corresponding rate, depend on the DNA conformation?  In
particular, is it important that the DNA is coiled at the length
scale larger than the persistence length?  Is it important that
DNA coil may not fit in the volume available, and then DNA must be
a globule, like in the nucleoid in a procaryotic cell \emph{in
vivo} or under experimental conditions \emph{in vitro}
\cite{Odijk}?  Second, closely related aspect is that BWH theory
\cite{BWH} does not answer the experimentally most relevant
question \cite{Halford} of the interplay between 1D sliding and 3D
diffusion.  In particular, one of the questions raised by
experiments and not answered by the BWH theory \cite{BWH} is about
the correlations between the place where a protein departs from
DNA and the place where it re-adsorbs.  Third aspect, although of
a lesser importance and more taste-dependent, BWH theory
\cite{BWH} does not yield simple intuitive explanation for
non-monotonic dependence of the rate on the strength of
non-specific adsorbtion, and one may want to know whether there
exists simple qualitative description of the rate at least in some
limits.

More recent refinement of the theory is given in the work Ref.
\cite{FrenchGroup}.  The authors of this work follow BWH in that
they treat DNA in terms of ``domains'' - a concept having no
unambiguous definition in the physics of DNA.  Also, the paper
Ref. \cite{FrenchGroup} makes it very explicit that BWH \cite{BWH}
and subsequent theories neglect correlations between the place
where protein desorbs from DNA and the place where it adsorbs
again - the approximation that clearly defies the polymeric nature
and fractal properties of DNA. At the same time, this
approximation leaves unanswered the experimentally motivated
question of the interplay between 1D and 3D components of the
search process.

In the recent years, the problem was revisited by physicists
several times \cite{Bruinsma,Marko,Mirny}, but the disturbing fact
was that all of them attributed quite different results and
statements to BWH: the paper Ref. \cite{Bruinsma} says that
according to BWH, the search time scales as DNA lengths $L$ rather
than $L^2$ as in 1D diffusion along DNA; the work Ref.
\cite{Marko} states that proteins slide along DNA some distance
which is independent of DNA conformation, regardless even of the
DNA fractal properties; the article Ref. \cite{Mirny}, although
concentrates on the role of the non-uniform DNA sequence, claims
that the time for 3D diffusion must be about the same as time for
1D diffusion along DNA. Further, possibly even more disturbing
fact is that neither of the papers
\cite{BWH,FrenchGroup,Bruinsma,Mirny} makes any clearly
articulated explicit assumption about DNA conformation. Is it
straight, or Gaussian coil with proper persistence length, or
what?  Does the result depend on the DNA conformation?
Interestingly, experimenters do discuss in their works (see
\cite{Halford} and references therein) the issue of correlated vs.
uncorrelated re-adsorbtion, these discussions call for theoretical
attention and theoretical description in terms of correlations in
fractal DNA, but so far proper theory was not suggested.

%To see immediately that DNA conformation is very important, we can
%offer the following simple argument.  Suppose DNA is a Gaussian
%coil, and assume for a second that its effective segment is unity.
%A protein diffusing along such DNA over some time $t$ moves along
%DNA by the contour distance which scales as $t^{1/2}$, but since
%DNA itself is a Gaussian coil, this corresponds to the
%displacement in space which is proportional only to $t^{1/4}$.
%This is very slow subdiffusion, it is clear that this is an
%extremely inefficient way of finding a far away target.  What is
%then the advantage of sliding along DNA?

Motivated by these considerations, we in this work set out to
re-examine the problem from the very beginning.  We explicitly
take into account that DNA is fairly straight at the length scale
smaller than persistence length, it is Gaussian coil on the larger
length scale.  We also consider the possibility that DNA is
confined within such a volume where Gaussian coil does not fit (as
it does not fit into a typical procaryotic cell, for instance), in
which case DNA must be a globule.

\subsection{Model, approach, and limitations}\label{sec:model}

We assume that within some volume $v$ some (double helical) DNA is
confined, with contour length $L$, persistence length $p$, and
with the target site of the size $b$.

We further assume that protein can be non-specifically adsorbed on
any place of the DNA, and that non-specific adsorbtion energy
$\epsilon$, or the corresponding constant $y = e^{\epsilon/k_BT}$,
is the same everywhere on the DNA and does not depend on the DNA
sequence.  We assume that every protein molecule has just one site
capable to adsorb on the DNA.  There are proteins with two such
sites, they can adsorb on two separate pieces of DNA at the same
time and thus serve as a cross-linker for the DNA itself.  We do
not consider this possibility in this article.

We assume that there is only one molecule of DNA.  In reality,
macroscopic sample of DNA solution at certain concentration is
used in any \emph{in vitro} experiment.  From the theoretical
standpoint, DNA solution with concentration of $1/v$ (in units of
DNA chains per unit volume) is equivalent to the system of one DNA
considered here.  We also assume that DNA has only one target site
on it, which is not always true in reality \cite{Halford}.

We assume that non-specifically bound protein can diffuse (slide)
along DNA with the diffusion coefficient $D_1$, while protein
dissolved in surrounding water diffuses in 3D with diffusion
constant $D_3$.  Thus, we have a unitless parameter related to the
diffusion coefficients, it is $d = D_1/D_3$. In the simpler
version of the theory, which we shall consider first, we assume
$D_1=D_3$, or $d=1$.  For simplicity, we assume that while protein
is diffusing, either in 3D or along the DNA, DNA itself remains
immobile.

The quantity of our interest is the time needed for the target
site to be found by a protein (consider. e.g., an example of
restriction enzyme attacking viral DNA intruder).  One should
imagine certain concentration $c$ of proteins randomly introduced
into the system, and ask what is the time needed for the
\emph{first} of these proteins to arrive to the target site. In
this paper, we will only address the mean time, averaged over both
thermal noise and DNA conformation. For this averaged quantity,
since the DNA is assumed immobile, the problem can be addressed in
a simple way, by looking at the stationary \emph{rate}.  Namely,
we should consider that there is a sink of proteins in the place
of the specific target site, and that it consumes proteins with
the rate $J$ proportional to concentration $c$, which should be
supported on a constant level by an influx to maintain
stationarity.  Obviously then, the averaged time is just $1/J$. At
the end of the paper, in section \ref{sec:single_protein_view} we
show how to re-derive all our results in terms of a single
protein, thus avoiding an artificial assumption that there is a
sink of proteins at the place of the target.

In this article, we calculate the rate $J$ assuming concentration
$c$ an arbitrary constant.  In order to compare the predicted rate
to the Smoluchowski rate $J_s = 4 \pi D_3 c b$, we shall mainly
look at the ratio
\be \frac{J}{J_s} = \frac{J}{4 \pi D_3 c b} \sim \frac{J}{D_3 c b}
\ , \ee
which characterizes the acceleration of the reaction rate achieved
due to the sliding along DNA.

We will be mainly interested in scaling dependence of the rate $J$
or acceleration $J/J_s$ on major system parameters, such as $y$,
$L$, and $v$. In this context, we will use symbol ``$\sim$'' to
mean ``equal up to a numerical coefficient of order one'', while
symbols $>$ and $<$ mean $\gg$ and $\ll$, respectively.

Along with dropping out all numerical coefficients in our scaling
estimates, we also make several assumptions driven by pure desire
to make formulae simpler and to clarify major physical ideas.  We
assume that all the ``microscopic'' length scales are of the same
order, namely, about target size $b$: protein diameter, double
helical DNA diameter, and the distance from DNA at which
non-specific adsorbtion takes place.  These assumptions are easy
to relax.

Throughout this work we disregard the excluded volume of DNA,
considering DNA coil as Gaussian and \emph{not} the swollen coil,
described by the Flory index $3/5$.  This is a reasonable
approximation for most realistic cases \cite{RedBook}.  Indeed,
for many real DNAs, such as, e.g., $\lambda$-DNA, it is justified
because of a large persistence length-to-diameter ratio of the
double helix: excluded volume in the coil remains unimportant up
to DNA length about $L < p^3/b^2$ (up to about 100000 base pairs
under normal non-exotic ionic conditions).  We further assume that
the volume fraction of DNA inside volume $v$, which is about
$Lb^2/v$, is sufficiently small even when DNA is a globule. In
particular, we assume $Lb^2/v < b/p$, because in a denser system
liquid crystalline nematic ordering of DNA segments becomes likely
\cite{RedBook}. Of course, real nucleoid is a rather complex
structure involving much more sophisticated features than just
orientational ordering, they are caused by structural and other
proteins, by entanglements, etc - see the recent experimental work
\cite{Odijk} and references therein.  In this paper we shall touch
neither of these issues, guided by the prejudice that simple
questions should be addressed first.

\subsection{Outline}

The plan of the article is as follows.  %After more detailed
%discussion of the model and underlying assumptions (section
%\ref{sec:model}),
In section \ref{sec:simple_case} we consider first the relatively
simple cases when DNA is a Gaussian coil and 1D sliding of
proteins along DNA involves
only a small part of DNA length. %(section \ref{sec:simple_case}).
Already in this situation we will be able to explain the effect of
correlated re-adsorbtion and arrive at a number of new results,
such as, for instance, possible asymmetric character of the
maximum on the curve of the rate as a function of adsorbtion
strength.  These results are also derived through the
electrostatic analogy in the appendix
(\ref{sec:electrostatic_analogy}).  In the section
\ref{sec:summary} we present a summary of all possible scaling
regimes.  We then discuss them in more details (section
\ref{sec:more_cases}).  We start this by looking at the rate
saturation when 1D sliding involves entire DNA length (section
\ref{sec:saturation}).  We then consider a delicate case when DNA
as a whole is a globule (section \ref{sec:globule}); in this case,
we found that even the 3D transport of proteins is in many cases
realized through the sliding of adsorbed proteins along DNA and
using DNA as a network of 1D transport ways.  We continue in
section \ref{sec:d_neq_1} by looking at the situations when
diffusion coefficient of the proteins along DNA is either smaller
or larger than their diffusion coefficient in the surrounding bulk
water.  In section \ref{sec:single_protein_view} we re-derive all
our major results using the language of single protein search time
instead of a stationary process and flux.  Finally, we conclude
with comparison of our results to those of earlier works and the
discussion of possible further implications of our work (section
\ref{sec:discussion}).

\section{Simple case: straight antenna vs. Gaussian coil
antenna}\label{sec:simple_case}

\begin{figure*}
\centerline{\scalebox{0.8}{\includegraphics{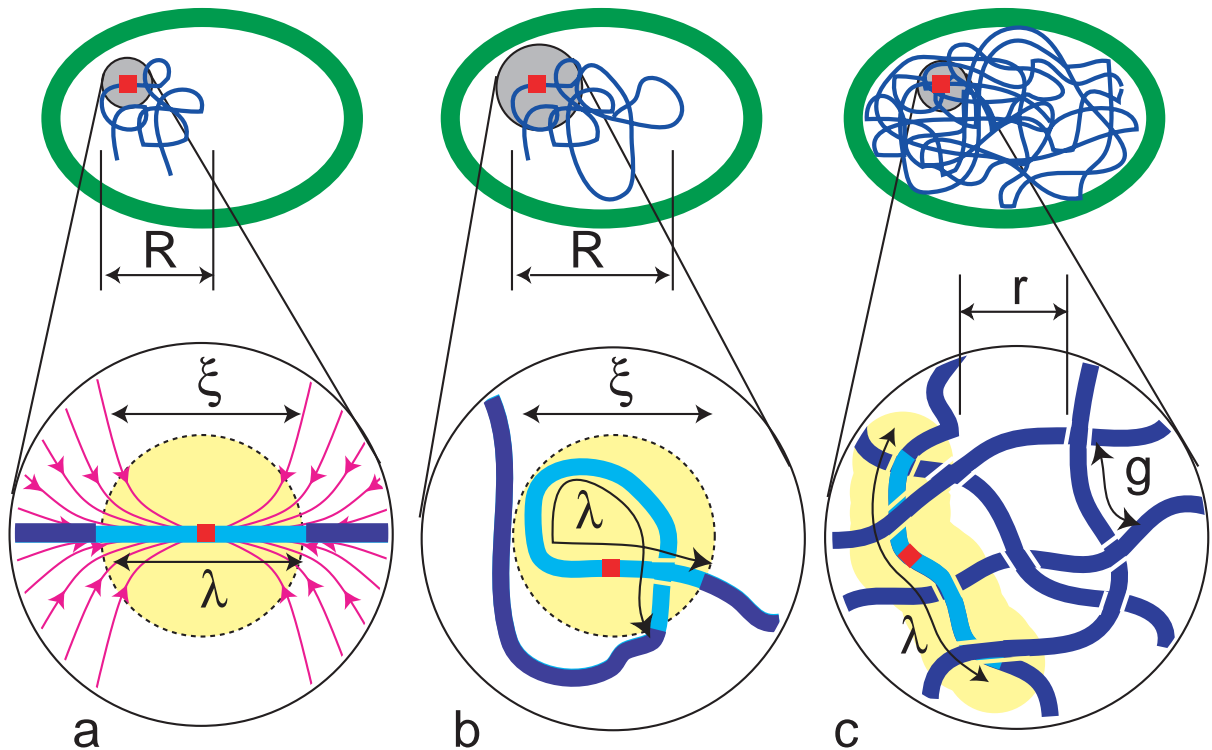}}}
%\centerline{\scalebox{0.8}{\includegraphics{antenna_tao2.pdf}}}
%\vspace{-15cm}
\caption{Antenna in a variety of cases.   The upper part of every
figure represents a poor man's idea of a prokaryotic cell. In
figures a and b, DNA in the cell is a coil, because coil size $R$
is smaller than the cell dimension; alternatively, one can think
of dilute solution of DNA in which $R$ is much smaller than the
distance to other coils (not shown). In figure c, the amount of
DNA is so large, that the coil size would have exceeded the cell
diameter, and so DNA is a globule; alternatively, one can think of
a semi-dilute solution \protect\cite{DeGennesBook} of strongly
overlapping DNA coils. The lower figures represent blow up view of
the region around the target site on DNA.  The antenna part of DNA
around the target is shown in lighter color than the rest of DNA.
The space region below the crossover length scale is shadowed.
This space region is roughly spherical in cases a and b, it is
sausage shaped in case c. Figure a also shows the averaged flow
lines of the diffusion, which go in 3D far away from the target
and go mostly along DNA within antenna length scale (they are
equivalent to electric field lines in terms of electrostatic
analogy, Appendix \protect\ref{sec:electrostatic_analogy}).  In
figures b and c flow lines are not shown, simply because it is
difficult to draw them.  In figure c, we see that DNA globule
locally looks like a temporal network, with the mesh size $r$. In
this case, antenna might be much longer that one mesh.  In the
figure, mesh size is not larger than persistence length, so the
length of DNA in the mesh $g$ is about the same as $r$; at lesser
density, mesh size might be longer, and then DNA in the mesh would
be wiggly, with $g \gg r$.} \label{fig:antenna}
\end{figure*}

The reason why non-specific adsorbtion on DNA can speed up the
finding of target is illustrated in Fig.   \ref{fig:antenna} (a)
and (b): it is because DNA forms a kind of an antenna around the
target thus increasing the size of the ``effective target''. How
should we determine the size of this antenna?  The simplest
argument is this.  Suppose antenna size is $\xi$ and contour
length of DNA inside antenna is $\lambda$. %(as with all lengths,
%we assume $\xi$ and $\lambda$ dimensionless, corresponding real
%lengths are $b \xi$ and $b \lambda$)
It is worth to emphasize that $\xi$ and $\lambda$ do not define
any sharp border, but rather a smooth cross-over, such that
transport outside antenna is \emph{mainly} due to the 3D
diffusion, while inside antenna transport is \emph{dominated} by
the sliding, or 1D diffusion along DNA. The advantage of thinking
about \emph{stationary} process is that under stationary
conditions, the flux of particles delivered by the 3D diffusion
into the $\xi$-sphere of antenna must be equal to the flux of
particles delivered by 1D diffusion into the target. The former
rate is given by the Smoluchowski formula (see appendix
\ref{sec:Smoluchowski}) for the target size $\xi$ and for the
concentration of ``free'' (not adsorbed) proteins $c_{\rm free}$,
it is $\sim D_3 c_{\rm free} \xi$. To estimate the latter rate, we
note that the time of 1D diffusion into the target site from a
distance of order $\lambda$ is about $\lambda^2/D_1$; therefore,
the rate can be written as $ \left( \lambda c_{\rm ads} \right) /
\left(\lambda^2 / D_1 \right)$, where $\lambda c_{\rm ads}$ is the
number of proteins non-specifically adsorbed on the piece of DNA
of the length $\lambda$.  Thus, our main \emph{balance} equation
for the rate $J$ reads
\be J \sim D_3 c_{\rm free} \xi  \sim \frac{D_1 c_{\rm ads} }{
\lambda } \ . \label{eq:balance} \ee
Formally, this equation follows from the continuity equation,
which says that divergence of flux must vanish everywhere for the
stationary process, flux must be a potential field.

Notice that the balance equation (\ref{eq:balance}) depends on the
relation between $\xi$ and $\lambda$ - between the size of antenna
measured in space ($\xi$) and measured along the DNA ($\lambda$).
Here, we already see why fractal properties of DNA conformations
enter our problem.

To determine the one-dimensional concentration of non-specifically
adsorbed proteins, $c_{\rm ads}$, and concentration of proteins
remaining free in solution $c_{\rm free}$, we now argue that as
long as antenna is only a small part of the DNA present, every
protein in the system will adsorb and desorb many times on DNA
before it locates the target, therefore, there is statistical
equilibrium between adsorbed and desorbed proteins. Assuming that
we know the adsorbtion energy $\epsilon$ or the corresponding
constant $y= e^{\epsilon / k_B T}$, and remembering that adsorbed
proteins are confined within distance or order $b$ from the DNA,
we can write down the equilibrium condition as
\be c_{\rm ads} / c_{\rm free} b^2 = y \ , \label{eq:equilibrium}
\ee
which must be complemented by the particle counting condition
\be c_{\rm ads} L + c_{\rm free} \left( v - Lb^2  \right) = cv \ .
\label{eq:particle_counting} \ee
Since volume fraction of DNA is always small, $Lb^2 \ll v$,
standard algebra then yields
\bea c_{\rm ads} & \simeq & \frac{cvyb^2}{yLb^2+v} \sim \left\{\begin{array}{lcr} cyb^2 & {\rm if} & y < v/Lb^2 \\
cv/L & {\rm if} & y > v/Lb^2
\end{array} \right. \ , \nonumber \\
c_{\rm free} & \simeq & \frac{cv}{yLb^2+v} \sim \left\{\begin{array}{lcr} c & {\rm if} & y < v/Lb^2 \\
cv/Lb^2y & {\rm if} & y > v/Lb^2
\end{array} \right. \ . \label{eq:equilibrium_concentrations} \eea
%

%\begin{figure*}
%\centerline{\scalebox{0.8}{\includegraphics{antenna_tao2.eps}}}
%\centerline{\scalebox{0.8}{\includegraphics{antenna_tao2.pdf}}}
%\vspace{-15cm}
%\caption{Antenna in a variety of cases.  Figure a depicts the
%situation when antenna is shorter than persistence length, and
%then antenna is straight.  This figure also shows the averaged
%flow lines of the diffusion, which go in 3D far away from the
%target and go mostly along DNA within antenna length scale. Figure
%b shows the situation when antenna is longer than persistence
%length and, therefore, antenna represents a Gaussian coil.  Figure
%c illustrates the case when DNA is a globule, so antenna consists
%of several blobs.  Let us stress, that particularly figure c
%presents only a very little piece of the entire globular DNA.}
%\label{fig:antenna}
%\end{figure*}

Note that at the length scales smaller than persistence length $p$
DNA double helix is practically straight, while on the length
scales greater than $p$, double helix as a whole is a Gaussian
coil. That means, if we take a piece of double helix of the
contour length $\lambda$, then its size in space scales as
\be \xi \sim \left\{\begin{array}{lcr} \lambda & {\rm when} & \lambda < p \\
\sqrt{\lambda p} & {\rm when} & \lambda > p
\end{array} \right. \ . \label{eq:fractality} \ee

Substituting this result into the balance equation
(\ref{eq:balance}), we can determine the antenna size and then,
automatically, the rate, the latter being either side of the
balance equation.  We have to be careful, because we see that
there are already as many as four different scaling regimes, due
to equations (\ref{eq:equilibrium_concentrations}) and
(\ref{eq:fractality}):
\begin{itemize} \item Regime A - antenna is straight (upper line
of Eq. (\ref{eq:fractality})), adsorbtion is relatively weak
(upper lines in the Eq. (\ref{eq:equilibrium_concentrations}));
\item Regime B - antenna is Gaussian (lower line in the Eq.
(\ref{eq:fractality}), but adsorbtion is still relatively weak;
\item Regime C - antenna is Gaussian and adsorbtion is relatively
strong (lower lines in the Eqs.
(\ref{eq:equilibrium_concentrations})); \item Regime D - Straight
antenna and strong adsorbtion. \end{itemize} Later we will find
plenty more regimes, but now let us consider just these ones, one
by one.

To begin with, suppose antenna is straight ($\lambda < p$, so
$\lambda \sim \xi$, see Fig.  \ref{fig:antenna}, (a)) and
non-specific adsorbtion relatively weak ($y<v/Lb^2$, so $c_{\rm
ads} \sim cyb^2$). In this case, balance equation yields $\lambda
\sim b(y d)^{1/2}$, or for the rate
\be J \sim c \sqrt{D_3D_1} y^{1/2} b \ ; \ee
in other words, for the ratio of this rate to the Smoluchowski
rate $J_s \sim D_3 c b$, we obtain
\be \frac{J}{J_s} \sim  (y d)^{1/2}  \ \ \ \ \  ({\rm regime \ A}).
\label{eq:straight} \ee
This result remains correct as long as antenna remains shorter
than persistence length, and since we know $\lambda$, we obtain
this condition explicitly:  $y < p^2/b^2 d$.

Let us now suppose that non-specific adsorbtion is still
relatively weak ($y<v/Lb^2$, so $c_{\rm ads} \sim cyb^2$), but it
is strong enough such that antenna is longer than persistence
length ($\lambda > p$, so that $\xi \sim \sqrt{\lambda p}$, see
Fig. \ref{fig:antenna}, (b)). Then our balance equation yields
$\lambda \sim \left( y d \right)^{2/3}p^{-1/3}b^{4/3}$ or
\be \frac{J}{J_s} \sim \left( \frac{y p d}{b} \right)^{1/3}  \ \ \
\ \ ({\rm regime \ B}). \label{eq:gaussian} \ee
One should check that this new result for $\lambda$ implies that
$\lambda > p$ at $y > p^2/b^2d$, and so $y \sim p^2 /b^2d$ is the
cross-over line between the two regimes, A and B. In both regimes,
and as expected, the rate grows with the strength of non-specific
adsorbtion, $y$, because increasing $y$ increases the size of
antenna.  However, the functional scaling dependence of the rate
on $y$ is significantly different, reflecting the difference in
DNA fractality at different length scales.

Before we proceed with analysis of other scaling regimes, it is
useful to make the following comment.  The balance equation
(\ref{eq:balance}) describes the fact that every protein going
through the 3D diffusion far away must then also go through the 1D
diffusion closer to the target.  In other words, balance equation
(\ref{eq:balance}) describes the self-establishing match between
3D and 1D parts of the process. But we can also look at the
situation differently: suppose that one particular protein is
adsorbed on DNA in a random place, and let us estimate the
distance it can diffuse along DNA before it desorbs due to a
thermal fluctuation. Since probability of thermally activated
desorbtion is proportional to $e^{-\epsilon/k_BT}=1/y$, the time
protein spends adsorbed must be about $b^2 y / D_3$. During this
time, protein diffuses along DNA by the distance about $\sqrt{D_1
b^2 y /D_3} = b \sqrt{yd}$.  Following \cite{FrenchGroup,Marko},
we call it \emph{sliding distance}.  We see, therefore, that
antenna length $\lambda$ is just about sliding distance for the
straight DNA, but $\lambda \gg \ell_{\rm slide}$ for the coiled
DNA.  This seems for the first glance like a very weird result:
how can possibly be antenna longer than the distance over which
protein can slide?  In fact antenna does become longer than the
bare sliding distance, and this happens because for the coiled DNA
every protein, desorbed after sliding the distance of the order of
$\ell_{\rm slide}$, has a significant chance to re-adsorb nearby.
Such correlated re-adsorbtion gets more likely as we consider more
and more crumpled conformations of DNA. Indeed, if we in general
assume that $\xi \sim \lambda^{\nu}$, then balance equation yields
$\lambda \sim y^{1/(1+\nu)}$, which means that $\lambda$ grows
with $y$ \emph{faster} than $\ell_{\rm slide} \sim y^{1/2}$ at
every $\nu < 1$.  This growth of $\lambda$ with $y$ gets
increasingly fast as $\nu$ decreases, which corresponds to more
crumpled conformations. We should emphasize that this mechanism of
correlated re-adsorbtion is impossible to see as long as DNA
polymeric and fractal properties are not considered explicitly,
that is why this mechanism has been overlooked in previous works.

With further increase of either non-specific adsorbtion strength
$y$ or DNA overall length $L$, we ran into the situation when most
of the proteins are adsorbed on the DNA.  In other words, if one
prefers to think in terms of a single protein diffusion, then this
single protein molecule spends most of the time adsorbed on DNA
far away from the target.  For this case, we have to use the lower
lines of the formulae (\ref{eq:equilibrium_concentrations}) and
substitute it into the balance equation (\ref{eq:balance}).  Since
equilibrium condition (\ref{eq:equilibrium}) is still satisfied,
the result $\lambda \xi \sim ydb^2$ remains unchanged.  Depending
on whether antenna length $\lambda$ is longer or shorter than
persistence length, we obtain the regimes C and D.

For regime C, we have $\lambda >p$, antenna is a Gaussian coil and
$\xi \sim \sqrt{\lambda p}$, yielding $\lambda \sim (y
d)^{2/3}p^{-1/3}b^{4/3}$ and
\be \frac{J}{J_s} \sim \frac{v (pd)^{1/3}}{L b^{7/3}y^{2/3}}  \ \
\ \ \ ({\rm regime \ C}). \label{eq:falling_rate} \ee
Given our expression for $\lambda$, the condition $\lambda > p$
implies the familiar $y>p^2/b^2d$, and another condition for this
regime is that most proteins are adsorbed, or $y > v/Lb^2$, see
Eqs. (\ref{eq:equilibrium_concentrations}).

For regime D, antenna is straight, so $\xi \sim \lambda$, and we
get $\lambda \sim b (yd)^{1/2}$, just as in the regime A.  For the
rate however substitution of lower lines of the Eqs.
(\ref{eq:equilibrium_concentrations}) into the balance equation
(\ref{eq:balance}) yields
\be \frac{J}{J_s} \sim \frac{v d^{1/2}}{L b^2 y^{1/2}}  \ \ \ \ \
({\rm regime \ D}). \label{eq:falling_rate_straight} \ee
According to our discussion, this regime should exist when $y
<p^2/b^2d$ and $y>v/Lb^2$.  As we shall see later, in the section
\ref{sec:d_neq_1}, these two conditions can be met together and
the room for this regime exists only if $d<1$, which means when 1D
diffusion along DNA is slower than 3D diffusion in space.

In both regimes C and D, overall rate decreases with the increase
of non-specific adsorbtion, $y$, because 3D transport to the
antenna is slowed down by the lack of free proteins.

We have so far discussed four of the scaling regimes, our results
are equations (\ref{eq:straight}), (\ref{eq:gaussian}),
(\ref{eq:falling_rate}) and (\ref{eq:falling_rate_straight}).
Already at this stage, we gained simple understanding of the
non-monotonic dependence of the rate on $y$ - phenomenon formally
predicted in \cite{BWH} and observed in \cite{BWH2}, but
previously not explained qualitatively: at the beginning,
increasing $y$ helps the process because it leads to increasing
antenna length; further increase of $y$ is detrimental for the
rate because it leads to an unproductive adsorbtion of most of the
proteins.  We have also obtained a new feature, absent in previous
works: the shape of the maximum on the $J(y)$ curve is asymmetric,
at least if DNA is not too long: in the regimes B and C, rate
grows as $y^{1/3}$ and then falls off as $y^{-2/3}$.

Since there are quite a few more scaling regimes, it is easier to
understand them if we now interrupt and offer the summary of all
regimes as presented in Figure \ref{fig:diagram_d_equal_1} and
Table \ref{tab:rates_table}.

\section{Summary of the results: scaling
regimes}\label{sec:summary}

\begin{table*}
\caption{The summary of rates and antenna lengths in various
regimes.  In labelling regimes, we skip J and L to avoid confusion
with the rate and DNA length.} \label{tab:rates_table}
\begin{tabular}{|l|l|l|l|}
\hline Regime & Description & $J/J_s$ & $\lambda$\\
\hline Axes & Smoluchowski: no antenna & $1$ & $b$\\
\hline A & straight antenna, few proteins adsorbed & $(yd)^{1/2}$
& $b(yd)^{1/2}$\\
\hline B & coiled antenna, few proteins adsorbed & $\left(ypd/b
\right)^{1/3}$ &
$ \left( yd \right)^{2/3} p^{-1/3}b^{4/3}$\\
\hline C & coiled antenna, most proteins adsorbed &
$\frac{v(pd)^{1/3}}{L b^{7/3} y^{2/3}}$ & $\left( yd \right)^{2/3} p^{-1/3} b^{4/3}$\\
\hline D ($d<1$) & straight antenna, most proteins adsorbed &
$\frac{vd^{1/2}}{L b^{2} y^{1/2}}$ & $b(yd)^{1/2}$\\
\hline E & whole DNA as straight antenna, few proteins adsorbed & $L/b$ & $L$\\
\hline F & whole DNA as coiled antenna, few proteins adsorbed & $\left( Lp /b^2 \right)^{1/2}$ & $L$\\
\hline G & whole DNA as antenna, most proteins adsorbed & $\frac{vd}{L^2b}$ & $L$\\
\hline H & antenna with coiled mesh, most proteins adsorbed & $\frac{p}{b^2} \left(\frac{vd}{Ly} \right)^{1/2}$ & $\frac{b}{p}\left(\frac{vyd}{L}\right)^{1/2}$\\
\hline I & antenna with straight mesh, most proteins adsorbed & $\frac{vd^{1/2}}{L b^{2} y^{1/2}}$ & $b \left( yd \right)^{1/2}$\\
\hline K ($d>1$) & antenna with straight mesh, few proteins adsorbed & $\left( yd \right)^{1/2}$ & $b \left( yd \right)^{1/2}$\\
\hline M ($d>1$) & antenna with coiled mesh, few proteins adsorbed & $p \left(\frac{Lyd}{v}\right)^{1/2}$ & $\frac{b}{p}\left(\frac{vyd}{L}\right)^{1/2}$\\
\hline
\end{tabular}
\end{table*}

\begin{figure}
\centerline{\scalebox{0.6}{\includegraphics{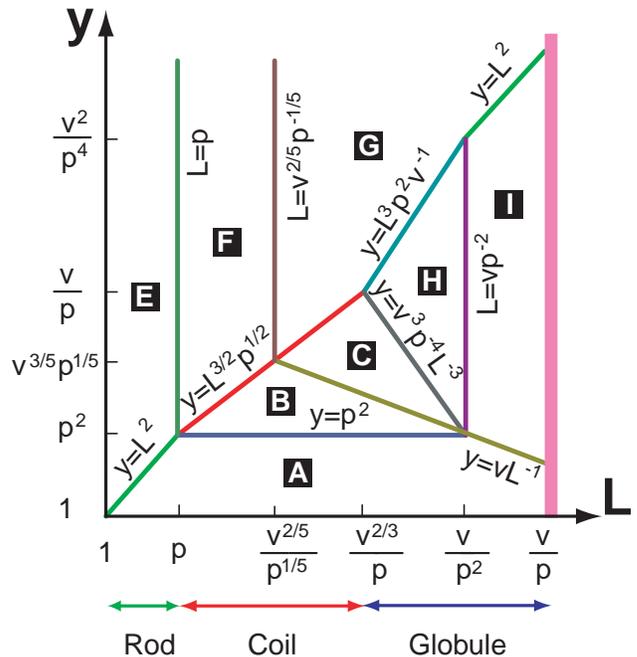}}}
\caption{Diagram of scaling regimes for the case $d=1$, when
diffusion along DNA has the same diffusion constant as diffusion
in surrounding water.  Both $L$ and $y$ axes are in the
logarithmic scale.  When DNA is shorter than persistence length
($b<L<p$) DNA is essentially a rod, DNA is a Gaussian coil as long
as it is longer than persistence length, but coil size is smaller
than the restriction volume $v$ ($p<L<v^{2/3}/p$), DNA is globular
at $L>v^{2/3}/p$, and we only consider $L$ up to about $v/pb$,
because at larger $L$ DNA segments start forming liquid
crystalline order.  Summary of the rates for each regime is found
in Table \protect\ref{tab:rates_table}.  Here, as well as in the
other figures, to make formulae look shorter, all lengths are
measured in the units of $b$, meaning that $L$, $p$, and $v$ stand
for $L/b$, $p/b$, and $v/b^3$.} \label{fig:diagram_d_equal_1}
\end{figure}

Our results are summarized in Fig. \ref{fig:diagram_d_equal_1} and
in the Table \ref{tab:rates_table}.  Figure
\ref{fig:diagram_d_equal_1} represents the log-log plane of
parameters $L$ and $y$, and each line on this plane marks a
cross-over between scaling regimes.  This figure gives the diagram
of scaling regimes for the specific case $d=1$ (or $D_1=D_3$);
later on, in the section \ref{sec:d_neq_1} we will return to the
more general situation and present corresponding diagrams for both
$d<1$ and $d>1$ cases.

To be systematic, let us start our review of scaling regimes from
the two trivial cases, which correspond to the axes in Fig.
\ref{fig:diagram_d_equal_1}. When $y \leq 1$, there is no
non-specific binding of proteins to the DNA, and no sliding along
DNA.  Proteins find their specific target at the rate which is
equal to the Smoluchowski rate, or $J/J_s =1$. Similarly, if the
DNA is very short, as short as the specific target site itself, or
$L \sim b$, then once again $J/J_s =1$ for trivial reason.  Since
we assume that there is some non-specific adsorbtion, or $y \geq
1$, and since DNA length is obviously always greater than the
target size $b$, our diagram in Fig. \ref{fig:diagram_d_equal_1}
presents only the $y>1$ and $L/b>1$ region, which is why pure
Smoluchowski regime is seen only on the axes.

If we increase $y$ and consider $y>1$ situation, then we have
significant non-specific adsorbtion of proteins on DNA, which
increases the rate due to the antenna effect.  If $y$ remains
moderate, the antenna is shorter than DNA persistence length, it
is straight.  This is regime labelled A in Fig.
\ref{fig:diagram_d_equal_1} and described by formula
(\ref{eq:straight}). With further increase of $y$, when $y >
p^2/b^2d$, we cross-over into the regime labelled B and described
by formula (\ref{eq:gaussian}), in this regime antenna is so long
that it is a Gaussian coil.   From the regime B, we can cross over
the line $y=v/Lb^2$ and get into the regime labelled C and
described by the formula (\ref{eq:falling_rate}).  One can
cross-over into the regime C by either increasing $y$ or
increasing $L$, because increasing either of these variables
promotes unproductive non-specific adsorbtion of proteins on far
away pieces of DNA and thus slows down the transport to the
specific target.

From regime A, we can also cross over the line $y=v/Lb^2$, but as
long as $d=1$ this does not bring us to the regime D, instead we
get to the new regime labelled I, which we will explain a few
lines below.

To understand all other scaling regimes, we have to remember that
our previous consideration throughout Section
\ref{sec:simple_case} was restricted in two respects.  First, we
assumed that the entire DNA in the form of Gaussian coil fits
within volume $v$, which is true only as long as $L < v^{1/3}$ and
$\sqrt{Lp} < v^{1/3}$, where $v^{1/3}$ stands for the linear
dimension of the restriction volume.  To relax this assumption, we
will have to consider a long DNA which is many times reflected by
the walls of volume $v$ and inside volume $v$ represents a
globule, locally looking like a semi-dilute solution of separate
DNA pieces, as illustrated in Fig. \ref{fig:antenna} (c). For such
long DNA, we shall find two more regimes labelled H and I in Fig.
\ref{fig:diagram_d_equal_1}. Second,  we assumed that the antenna
length $\lambda$ was smaller than full DNA length $L$; the
consequence of this was our statement (\ref{eq:equilibrium}) that
there is equilibrium between adsorbed and dissolved proteins.
Relaxing this assumption, we will have to discuss regimes labelled
E, F, and G on Fig. \ref{fig:diagram_d_equal_1}.

\begin{figure}
\centerline{\scalebox{0.4}{\includegraphics{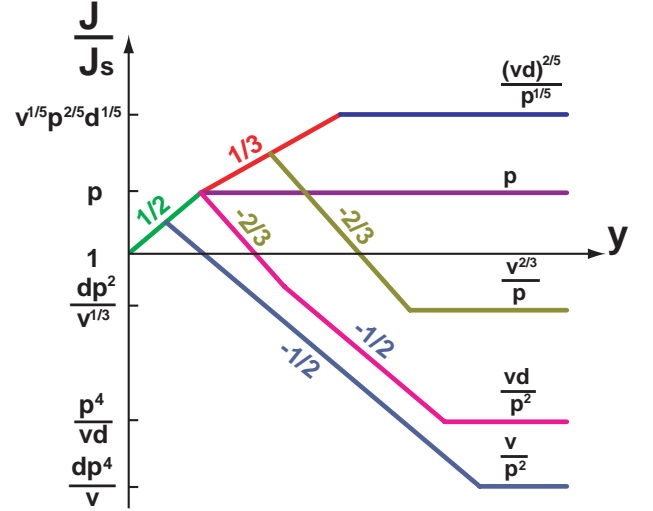}}}
\caption{Schematic representation of rate dependence on $y$. Both
the rate $J$ and $y$ are given in logarithmic scale.  The fraction
next to each curve shows its slope, which is the power of $J(y)$
dependence.  Each curve corresponds to the specified value of DNA
length $L$, also indicated in Figure
\protect\ref{fig:diagram_d_equal_1}, the length $L$ is shown above
the right end of each curve.  Experimentally, the value of $y$ can
be controlled through the salt concentration, because non-specific
adsorbtion of proteins is controlled by Coulomb interaction
between negative DNA and positive patch on the protein surface;
for instance, if the salt is ${\rm KCl}$, then it is believed
\cite{BWH2,Bruinsma} that $y = 10 \left[ {\rm KCl} \right] +2.5$,
where $\left[ {\rm KCl} \right]$ is the molar concentration of the
salt.  Note that we recover the possibility, first indicated in
\protect\cite{BWH}, that the rate goes through the maximum and
then saturates, but in our case maximum is in many cases
asymmetric, while at large $y$ the rate becomes very small $J/J_s
\ll 1$, particularly for long DNA.  Here, as well as in the other
figures, to make formulae look shorter, all lengths are measured
in the units of $b$, meaning that $L$, $p$, and $v$ stand for
$L/b$, $p/b$, and $v/b^3$.} \label{fig:rate}
\end{figure}

In Figure \ref{fig:rate}, we present a schematic $y$-dependence of
the rate for a number of values of DNA lengths $L$.  Each curve is
labelled with the corresponding value of $L$.  To be specific, we
have chosen the lengths which correspond to various cross-overs
and are marked on the scaling regimes diagram, Figure
\ref{fig:diagram_d_equal_1}. Note that in many cases our result
for the rate exhibits a maximum and saturation beyond the maximum
- features first described in the work BWH, Ref. \cite{BWH}.
Unlike BWH, we find that the maximum is asymmetric and, even more
importantly, $J/J_s$ can become much smaller than unity, i.e., one
can observe deceleration in comparison with Smoluchowski rate. We
also find a number of other features, such as specific power law
scaling behavior of the rate.

Thus, we have to discuss one by one all the new regimes E, F, G,
H, I. This is what we do in the next section \ref{sec:more_cases}.

\section{Systematic consideration of scaling
regimes}\label{sec:more_cases}

\subsection{DNA is not long enough for full
antenna}\label{sec:saturation}

If DNA is too short for antenna, then proteins already adsorbed on
DNA can find their target faster than new proteins can be
delivered to the DNA from solution.  There is no adsorbtion
equilibrium any longer, and instead of formula
(\ref{eq:equilibrium}) we can only claim that $c_{\rm ads} < y
c_{\rm free}b^2$.  Therefore, the amount of adsorbed proteins
under stationary conditions is physically determined by the
stationarity itself, which means, we have to look at formula
(\ref{eq:balance}) as \emph{two} equations.  In doing so, we have
to replace $\lambda$ in the right hand side (one-dimensional rate)
by $L$, because we don't have more DNA than $L$, and we have to
replace $\xi$ in the left hand side, which is the antenna size for
3D transport, by $R$ - overall size of DNA coil.  Of course,
particle counting equation (\ref{eq:particle_counting}) is still
valid, it is the third equation.  Thus, our equations read:
\bea && \frac{J}{J_s}  \sim  \frac{c_{\rm free } R}{cb} \ ;
\nonumber \\
&& c_{\rm free} R  \sim  \frac{c_{\rm ads} d}{L} \ ; \nonumber \\
&& c_{\rm ads} L + c_{\rm free} v  \sim  cv \ .
\label{eq:three_equations} \eea
From here, we find
\be \frac{J}{J_s} = \frac{vRd/b}{RL^2+vd} \ . \label{eq:long_DNA}
\ee
We can now easily address all possible scaling regimes in which
antenna is longer than DNA.

To begin with, it is possible that DNA length is shorter than DNA
persistence length $L<p$, such that the entire DNA is essentially
straight, and then $R \simeq L$.  Assuming also $L^3 < v$, we
arrive at the scaling regime labelled E in Fig.
\ref{fig:diagram_d_equal_1}, in this regime
\be \frac{J}{J_s} \sim \frac{L}{b} \ \ \ \ \  ({\rm regime \ E}).
\ee
The borderline of this regime can be established from the
condition that since entire DNA is smaller than ``equilibrium''
antenna, we must expect that $c_{\rm ads}$ is smaller than its
equilibrium value, or $c_{\rm ads}/c_{\rm free}b^2 \leq y$.  Since
according to the second of the formulae (\ref{eq:three_equations})
we have $c_{\rm ads}/c_{\rm free} = LR/d$, so we have the
condition $LR/d < y b^2$; at $L<p$ this yields $y>L^2/b^2d$.  At
the same condition we can also arrive from the other side of the
crossover, by noting that regime A continues as long as antenna is
shorter than entire DNA, $\lambda < L$; using our result for
$\lambda$ for the regime A, this produces the same cross-over line
between regimes A and E.

For longer DNA, when $L>p$, entire DNA is Gaussian coil, its size
is $R \sim (Lp)^{1/2}$.  Still assuming that the second term
dominates in the denominator in formula (\ref{eq:long_DNA}), we
arrive at
\be \frac{J}{J_s} \sim \left( \frac{L p}{b^2} \right)^{1/2} \ \ \
\ \ ({\rm regime \ F}). \ee
This regime is labelled F in Fig.  \ref{fig:diagram_d_equal_1}.
Its borderline with regime E is obviously vertical line $L=p$.  As
regards cross-over to the regime B, once again it can be
established either from $c_{\rm ads}/c_{\rm free} = LR/d <y$ for
the regime F or from $\lambda < L$ for the regime B.  In either
way we arrive at the cross-over condition $y=L^{3/2}p^{1/2}/b^2
d$.

For even longer DNA, the antenna length becomes equal to the
length of entire DNA only at so large $y$, that the system is
already in the regime C, with rate falling down with increasing
$y$ because of the unproductive adsorbtion of proteins.  Since
antenna length $\lambda$ in the regime C is given by the same
formula as in the regime B, so the upper border line of the regime
C is the continuation of the corresponding line bordering regime
B, it is $y=L^{3/2}p^{1/2}/b^2d$.  However, when we cross this
line upwards from the regime C, we arrive at the new situation,
because now the first term dominates in the denominator of the
equation (\ref{eq:long_DNA}), meaning that most of the proteins
are adsorbed on DNA, such that we obtain
\be \frac{J}{J_s} \sim \frac{vd}{L^2b} \ \ \ \ \  ({\rm regime \
G}). \ee
The cross-over between this regime and regime F is vertical line
at which both terms are comparable in the denominator of equation
(\ref{eq:long_DNA}), it is $L = (vd)^{2/5}/p^{1/5}$.  Crossover
line with the regime C can once again be established from the
condition $c_{\rm ads}/c_{\rm free} = LR/d <y$.

In all regimes E, F, and G the rate saturates with increasing $y$.
For the regimes E and F this happens after just initial growth of
rate; for the regime G saturation occurs after rate goes through
the maximum and starts decreasing.  In all cases saturation is due
to the fact that increasing adsorbtion strength does not lead to
any increase of the antenna size, because already the entire DNA
is employed as antenna and antenna has nowhere to grow.

\subsection{Cell is not big enough to house DNA Gaussian
coil}\label{sec:globule}

When DNA is very long for a given volume, specifically, when
$(Lp)^{1/2} > v^{1/3}$, DNA cannot remain just a coil, it must be
a globule, as it is forced to return many times back into the
volume after touching the walls (see, for instance, \cite{RedBook}).
%We already mentioned in the
%Introduction, that even for a prokaryotic cell DNA packed in a
%nucleoid in quite a complex manner \cite{Odijk}, but here we
%resort for simplicity to the model well known in polymer physics
%(see, for instance, \cite{RedBook}) when DNA is just a
%structureless polymeric globule.
For the purposes of this work, it is sufficient to keep assuming
that excluded volume of DNA is not important, because volume
fraction of DNA within confinement volume $v$ is still small, and
even small compared at $b/p$. Nevertheless, the system locally
looks like a so-called semi-dilute solution of DNA, or transient
network with certain mesh size (see Figure \ref{fig:antenna}c).
%The major new aspect of this situation for our problem in this
%paper is the fact that even the 3D transport of proteins can be
%now be realized via non-specific adsorbtion on DNA and sliding
%along DNA, provided that protein can switch from time to time from
%one piece of DNA ``network'' to another.

%This mechanism of 3D transport  in a dense polymer system has
%potential interest far beyond present problem, and we plan to
%publish elsewhere its detailed analysis.
%This mechanism of 3D
%transport must inevitably become dominant at large enough $y$ and
%$L$, because, as we saw already for the regime D, at large $y$ the
%solution is depleted of freely dissolved proteins which suppresses
%the overall rate.

We should remind some basic facts regarding the semi-dilute
solution of transient network \cite{DeGennesBook,RedBook}.  Let us
denote $r$ the characteristic length scale of a mesh in the
network, it is in the scaling sense the same as the characteristic
radius of density-density correlation (see Figure
\ref{fig:antenna}c).  Let us further denote $g$ the characteristic
length along the polymer corresponding to the spatial distance
$r$. Quantities $r$ and $g$ can be estimated from the following
physical argument \cite{DeGennesBook,RedBook}. Consider a piece of
polymer of the length $g$ starting from some particular monomer,
it occupies region $\sim r^3$ and makes density about $\sim
g/r^3$; this density must be about overall average density, which
for our system is of the order of $L/v$. Thus, $g/r^3 \sim L/v$.
Second relation between $g$ and $r$ is similar to formula
(\ref{eq:fractality}), it depends on whether mesh size is bigger
or smaller than persistence length $p$:
\be r \sim \left\{ \begin{array}{lcr} g & {\rm if} & g < p \\
\sqrt{gp} & {\rm if} & g >p \end{array} \right. \ . \ee
Accordingly, we obtain after some algebra
\be \begin{array}{lccr} g \sim \sqrt{\frac{v}{L}} \ , &
r \sim \sqrt{\frac{v}{L}} & {\rm if} & L > \frac{v}{ p^2} \\
g \sim \frac{v^2}{L^{2}p^{3}} \ , & r \sim \frac{v}{Lp} & {\rm if}
& \frac{v^{2/3}}{p} < L < \frac{v}{p^2}
\end{array}  \ . \label{eq:blob_size} \ee
The upper line corresponds to the network so dense that every mesh
is shorter than persistence length and polymer is essentially
straight within each mesh.  The lower line describes much less
concentrated network, in which every mesh is represented by a
little Gaussian coil.

Returning to our problem, we should realize that the antenna
length $\lambda$ can in fact be longer than the mesh size $g$, as
illustrated in Fig.  \ref{fig:antenna} (c).  To estimate the
antenna size for this case, we should remember that desorbtion
from antenna does not necessarily completely breaks the sliding
along DNA, because protein can still re-adsorb on a nearby place
of DNA, more generally - on a \emph{correlated} place on DNA. To
account for this, let us imagine that the antenna part of DNA is
decorated by a tube of the radius $r$.  Since $r$ is the
correlation length in the DNA solution, protein remains correlated
with antenna as long as it remains within this tube around
antenna.  Accordingly, our main balance equation
(\ref{eq:balance}) must be modified to account for the fact that
3D transport on the scales larger than $r$ is now realized through
DNA network and, therefore, the task of regular 3D diffusion is
only to deliver proteins over the length scale of order of one
mesh size $r$, into any one of the $\lambda / g$ network meshes
along the antenna. The rate of delivery into one such mesh would
be $\sim D_3 c_{\rm free} r$, so overall delivery rate into the
antenna tube scales as $\sim D_3 c_{\rm free} r \lambda / g$.  As
usual, this must be equal to the rate of 1D delivery along antenna
into the specific target, so instead of (\ref{eq:balance}) we
finally get
\be J \sim D_3 c_{\rm free} r \frac{\lambda }{ g} \sim D_1
\frac{c_{\rm ads}}{\lambda} \ . \label{eq:balance_mesh} \ee
As long as antenna is shorter than the entire DNA, the relation
between $c_{\rm free}$ and $c_{\rm ads}$ equilibrates and obeys
(\ref{eq:equilibrium}-\ref{eq:equilibrium_concentrations}), so we
finally get
\be  \lambda^2 \sim b^2\frac{g  y d}{r} \ , \ee
and
\be \frac{J}{J_s} \sim \frac{c_{\rm free}}{c} \frac{r \lambda}{g}
\sim \frac{v}{Lb^2}\left( \frac{r  d}{y g} \right)^{1/2} \ . \ee
What is nice about this formula is that it remains correct in a
variety of circumstances - when antenna is straight ($\lambda
<p$), or antenna is Gaussian ($p< \lambda < v^{2/3}/p$), or
antenna is a globule ($\lambda > v^{2/3}/p$).

Taking $r$ and $g$ from the formulae (\ref{eq:blob_size}), we
finally obtain two new regimes.  When every mesh is Gaussian,
\be \frac{J}{J_s} \sim \frac{p}{b^2} \left( \frac{v d}{Ly}
\right)^{1/2} \ \ \ \ \  ({\rm regime \ H}).
\label{eq:Gaussian_mesh}\ee
This regime borders regime C along the line where antenna size is
equal to the mesh size, $\lambda = g$, which reads $y =
v^3/(L^3p^4b^2d)$.  Regime H also borders regime G along the line
where antenna size is as long as the entire DNA, $\lambda = L$, or
$y = L^3p^2/vb^2d$.  Finally, regime H also borders another regime
I along the vertical line $L = v/p^2$, which corresponds to DNA
within every mesh becoming straight (shorter than persistence
length).  For this regime, we have to use upper line in formulae
(\ref{eq:blob_size}), thus obtaining
\be \frac{J}{J_s} \sim  \frac{v d^{1/2}}{Lb^2y^{1/2}} \ \ \ \ \
({\rm regime \ I}). \label{eq:straight_mesh} \ee
This regime borders saturation regime G along the line
$y=L^2/b^2d$ where $\lambda = L$.

As regards the lower border of the regime I, it corresponds to the
situation when antenna becomes straight, which happens at
$y=v/Lb^2d$.  However, as long as $d=1$, which is the case
presented in Figure \ref{fig:diagram_d_equal_1}, this line
coincides with the line $y=v/Lb^2$ below which most proteins are
desorbed and free in solution.  That is why at $d=1$, there is no
room for the regime D, in which antenna is straight, but most
proteins adsorbed.  Indeed, when $d=1$, then 3D transport is
mostly realized by sliding along the network edges as soon as most
proteins are adsorbed, which precisely means that regime A crosses
over directly to regime I.

As we see, in both H and I regimes the rate $J$ decreases with
growing $y$, but does so slower than in the regime C, only as
$y^{-1/2}$ instead of $y^{-2/3}$.  This happens because adsorbed
proteins are not just taken away from the process, as in the
regime C, but they participate in 3D transport through the
network, albeit this transport is still pretty slow.

This completes our scaling analysis for the $d=1$ case shown in
Fig. \ref{fig:diagram_d_equal_1}.

\subsection{Diffusion rate along DNA is different from that in surrounding water}\label{sec:d_neq_1}

Let us now relax the $d=1$ condition and examine the cases when
diffusion along DNA is either slower ($d<1$) or faster ($d>1$)
than in surrounding water.

\begin{figure*}
\centerline{\scalebox{0.6}{\includegraphics{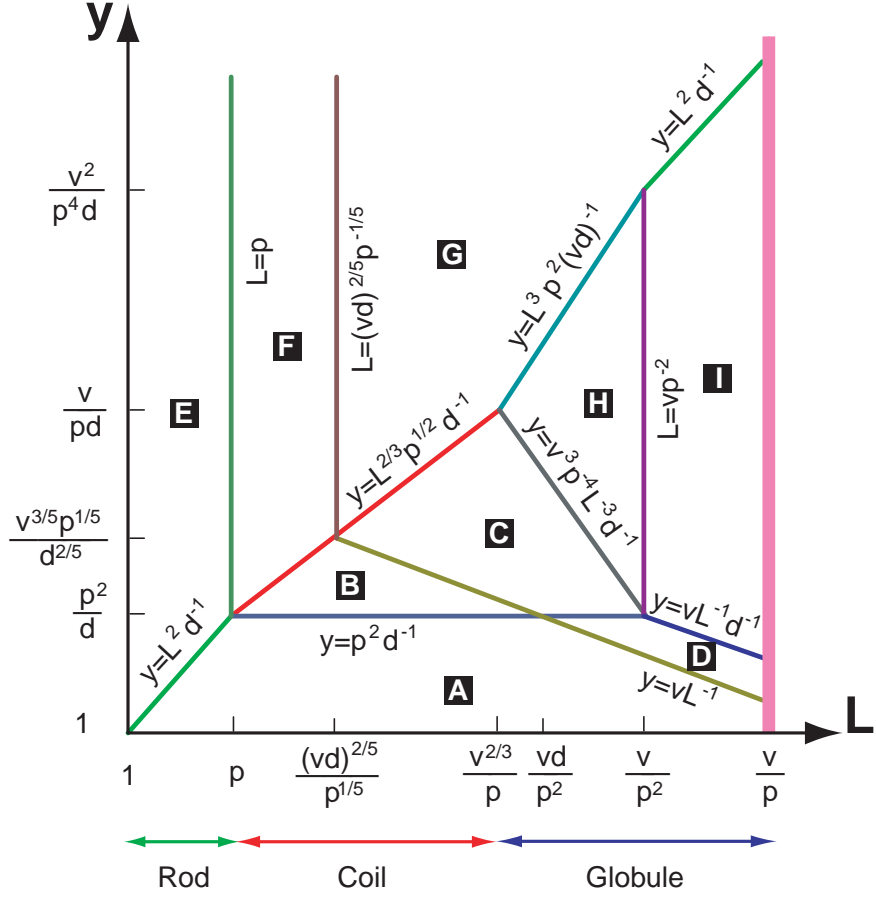}}}
\caption{Scaling regimes for the case $d<1$.  The major difference
from the $d=1$ case is the presence of regime D, in which majority
of proteins are adsorbed, but still the dominant 3D transport is
the usual diffusion through the surrounding water, because sliding
along DNA is too slow ($D_1<D_3$).  In this figure, as well as in
the other figures, to make formulae look shorter, all lengths are
measured in the units of $b$, meaning that $L$, $p$, and $v$ stand
for $L/b$, $p/b$, and $v/b^3$.} \label{fig:diagram_d_smaller_1}
\end{figure*}

First let us consider $d<1$ case, when diffusion along DNA is
slower than that in the surrounding water ($D_1<D_3$),
corresponding scaling regimes are summarized in the diagram Figure
\ref{fig:diagram_d_smaller_1}.  Most of the diagram is
topologically similar to that in the Figure
\ref{fig:diagram_d_equal_1}, and we do not repeat corresponding
analysis.  Of course, there are now powers of $d$ in all
equations, but the major qualitative novelty is that there is now
a room for the regime D sandwiched between regimes A and I. The
formal reason why this regime now exists in a separate region is
because the line $y=v/Lb^2d$ goes above the line $y=v/Lb^2$. To
understand the more meaningful physical difference, let us recall
that the line $y=v/Lb^2$ marks the cross-over above which most of
the proteins are adsorbed, but it is not enough for the
sliding-along-network
mechanism to dominate in the 3D transport at $d<1$. %, because even
%if most proteins are adsorbed, their sliding along DNA is so slow
%($d<1$) that the regular 3D diffusion of free proteins through
%water still dominates.  The collective diffusion takes over only
%when $y$ exceeds the value of $v/Lb^2d$.

Interestingly, the rate for both regimes D and I is given by the
same formula - compare Eqs. (\ref{eq:falling_rate_straight}) and
(\ref{eq:straight_mesh}).  This happens because antenna is
straight for the regime D and, while antenna is not straight for
the regime I, it still consists of a number of essentially
straight pieces, each representing one mesh.  The major difference
between regimes D and I, despite similar scaling of the rate, is
in the mechanism of diffusion: in the regime D, proteins diffuse
through the water in a usual manner, while in the regime I they
are mostly transported along the network of DNA, with only short
``switches'' on the scale of one mesh size $r$ between sliding
tours.  This is why straight pieces of DNA in different meshes
independently add together to yield the same overall formula for
rate as in the regime D.

\begin{figure*}
\centerline{\scalebox{0.6}{\includegraphics{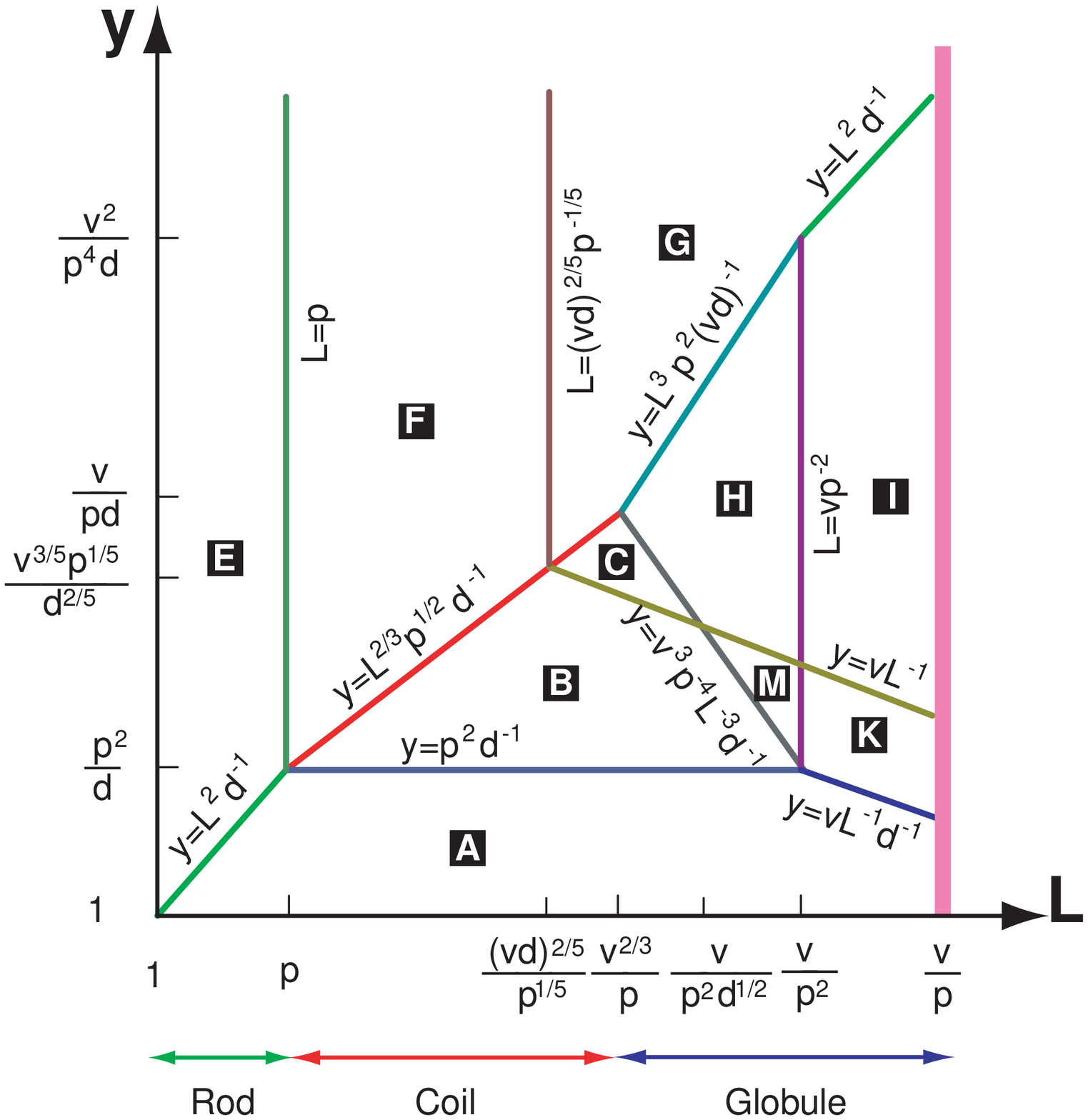}}}
\caption{Scaling regimes for the case $d>1$.  The major new
feature of this diagram compared to previous ones is the presence
of regimes K and M.  In these regimes the majority of proteins are
not adsorbed, but still the dominant 3D transport mechanism is the
sliding of minority proteins along DNA network, because it is so
much faster ($D_1>D_3$). We skip J and L in labelling regimes to
avoid confusion with rate $J$ and DNA length $L$.  In this figure,
as well as in the other figures, to make formulae look shorter,
all lengths are measured in the units of $b$, meaning that $L$,
$p$, and $v$ stand for $L/b$, $p/b$, and $v/b^3$.}
\label{fig:diagram_d_larger_1}
\end{figure*}

Let us now switch to the opposite limit and consider the $d>1$
case, for which the results are summarized in Figure
\ref{fig:diagram_d_larger_1}.  This diagram is quite similar to
the previously considered ones in Figures
\ref{fig:diagram_d_equal_1} and \ref{fig:diagram_d_smaller_1},
except there are now two new regimes labelled K and M (in
alphabetical labelling of the regimes we skip J and L to avoid
confusion with rate and DNA length).  These regimes are both below
the line $y=v/Lb^2$, which means, most of the proteins are not
adsorbed. However, since $d>1$, the new physical feature of the
situation is that adsorbed proteins, although they are in
minority, can nevertheless dominate in 3D transport by sliding
along DNA network, because sliding is now so fast at $d>1$.  Thus,
regimes K and M are the ones in which effective diffusion along
DNA network dominates, so we have to use formula
(\ref{eq:balance_mesh}) for the rate and antenna size, while for
the concentrations of free and adsorbed proteins we have to use
upper lines in the formulae (\ref{eq:equilibrium_concentrations}).
In the regime K, local concentration of DNA segments is so high,
that every mesh in DNA network contains an essentially straight
piece of DNA, so we have to use the upper line in formula
(\ref{eq:blob_size}), yielding (after some algebra)
\be \frac{J}{J_s} \sim  (yd)^{1/2} \ \ \ \ \ ({\rm regime \ K}).
\label{eq:straigh_mesh_large_d} \ee
Similarly, in the regime M mesh of the DNA network is Gaussian, we
have to use lower line in equation (\ref{eq:blob_size}), and this
produces
\be \frac{J}{J_s} \sim  p \left( \frac{L y d}{v} \right)^{1/2} \ \
\ \ \ ({\rm regime \ M}). \label{eq:Gauss_mesh_large_d}\ee
Since the majority of proteins are not adsorbed, it is not
surprising that rate grows with $y$ in both regimes K and M.
Notice that the rate is given by the same formula for the regimes
A and K - compare (\ref{eq:straight}) and
(\ref{eq:straigh_mesh_large_d}). This is similar to the situation
with regimes D and I, as discussed before, because although rate
is given by the same formula, the underlying diffusion mechanism
is fundamentally different.  In both cases of D and I or A and K,
it is possible that although scaling laws are the same, the
numerical pre-factors are different.

It is also interesting to note that the cross-over between regimes
B and M takes place on the line $y=v^3/p^4L^3b^2d$ where antenna
length is equal to the DNA length in one mesh: on the side of B
regime, antenna is shorter than one mesh, and transport to antenna
must be through water; on the side of M, antenna is longer than
one mesh, and effective transport along DNA network is at play.

\subsection{Maximal rate}\label{sec:maximal_rate}

To finalize our discussion of scaling regimes, it is reasonable to
ask: what is the maximal possible rate?  According to our results,
the maximal rate is achieved on the border between regimes F and
G, that is, at $L \sim (vd)^{2/5}/p^{1/5}$ and at $y \geq
v^{3/5}p^{1/5}/b^2d^{2/5}$.  Maximal possible acceleration
compared to Smoluchowski rate is about $(v p^2 d/b^5)^{1/5}$.  It
is interesting to note that the ``optimal strategy'' in achieving
the maximal rate at the minimal possible $y$ requires to have the
adsorbtion strength $y$ right at the level at which the
probability of non-specific adsorbtion for every protein is about
$1/2$ (on the line $y \sim v/L$).

It is interesting that the maximal possible acceleration grows
with overall volume $v$, which may seem counterintuitive.  This
result is due to the fact that total amount of DNA grows with
increasing $v$, and, according to our assumption, all this DNA has
still just one target.

\section{Discussion}\label{sec:discussion}

\subsection{Single protein view}\label{sec:single_protein_view}

Many of the previous theoretical works
\cite{FrenchGroup,Bruinsma,Mirny,Marko} looked at the situation in
terms of a single protein molecule diffusing to its target.  In
this view, one should imagine that a protein molecule is initially
introduced into a random place within volume $v$, and then one
should ask what is the first passage time \cite{Redner} needed for
the protein to arrive to the specific target site on DNA.  The
mean first passage time $\tau$ can of course be found using our
results for the rate $J$ by inverting the value of the rate and
assuming that on average there is just one protein molecule in the
system at any time: $\left. \tau  = 1/J\right|_{c=1/v} $. However,
we want to re-derive all our results directly in terms of $\tau$
in order to build bridges to the works of other authors.  The
re-derivation turns out also quite illuminating.

First let us consider that DNA is a globule, $L >v^{2/3}/p$ (or
semi-dilute solution), and look at the regimes H, I, K, and M;
unlike stationary diffusion approach above, in the single protein
language the derivation for the globular DNA case is actually
simpler. Following \cite{Mirny}, we imagine that the search
process for the given single protein consists of tours of 1D
sliding along DNA followed by diffusion in 3D, followed by 1D
sliding, etc.  If in one tour of 1D sliding protein moves some
distance $\lambda$ along DNA, then it takes time about
$\lambda^2/D_1$.  The length $\lambda$ here is, of course, our
familiar antenna length, but we will re-derive it here, so we
\emph{do not} assume it known.  As regards the tour of 3D
diffusion, it breaks correlation of the 1D sliding if it carries
protein over a distance larger or about the correlation length in
the DNA system, which is $r$ - mesh (or blob) size. Thus, the
longevity of one tour of 3D diffusion is about $r^2/D_3$.

The next step of our argument is this.  On its way to the target,
the protein will go through great many adsorbtion and de-sorbtion
cycles, therefore, the ratio of times protein spends adsorbed and
de-sorbed should simply follow equilibrium Boltzmann statistics:
\be \frac{\lambda^2/D_1}{r^2/D_3} \sim \frac{yLb^2}{v} \ .
\label{eq:ratio_of_times_in_globule} \ee
(Here, we note parenthetically that there is an approximation
underlying our argument: one tour of ``correlated 1D sliding''
does include small 3D excursions of the protein into water, but
they are small in the sense that they do not go beyond the
cross-over correlation distance and, therefore, re-adsorbtion
after excursion occurs on a correlated place on DNA. Accordingly,
these excursions make only marginal contribution to the sliding
time which is correctly estimated as $\sim \lambda^2/D_1$.)

The final part of the argument is most clearly formulated by
Bruinsma in the work ref. \cite{Bruinsma}: since subsequent tours
of 1D sliding occur over uncorrelated parts of DNA, full search
requires about $L/\lambda$ rounds.  Therefore, the total search
time $\tau$ can be written as
\be \tau \sim \frac{L}{\lambda} \left[ \frac{\lambda^2}{D_1} +
\frac{r^2}{D_3} \right] \ . \label{eq:full_time_for_globule} \ee

Equations (\ref{eq:ratio_of_times_in_globule}) and
(\ref{eq:full_time_for_globule}) solve the problem for all regimes
of globular DNA if we remember that mesh (or blob) size $r$ is
given by the formula (\ref{eq:blob_size}).  Notice that formula
(\ref{eq:ratio_of_times_in_globule}) gives a new interpretation to
the line $y \sim v/Lb^2$ on any of our diagrams Fig.
\ref{fig:diagram_d_equal_1}, \ref{fig:diagram_d_smaller_1},
\ref{fig:diagram_d_larger_1}: for the parameters below this line
most of the overall search time is spent in 3D diffusion, while
for the system with parameters above the line the major time
consuming part is 1D sliding.  It is close to this line where the
result of the work ref. \cite{Mirny} applies and these two times
are of the same order.  And let us remind that it is also close to
this line where the maximal possible rate is achieved (see section
\ref{sec:maximal_rate}).

Thus, four regimes H, I, K, and M result from two possibilities
for $r$ in Eq. (\ref{eq:blob_size}) (straight or Gaussian DNA
within a mesh) and two possibilities of either first or second
term dominance in formula (\ref{eq:full_time_for_globule}).

Let us now turn to the regimes A, B, C, and D, when DNA is a coil.
In this case, we still essentially rely on the equations similar
to (\ref{eq:ratio_of_times_in_globule}) and
(\ref{eq:full_time_for_globule}), except some effort is now needed
to understand the time of 3D diffusion.
%, or, in other words, what
%plays the role of $r$ in these two formulae.
%To begin with, let us note that the situation can be viewed in two
%equivalent ways. Either we have a single coil of DNA confined
%within volume $v$, or, equivalently but perhaps more appropriately
%for an \emph{in vitro} experiment, we have a dilute solution of
%DNA coils, each having one target site. In this view, we should
%assume concentration of $1/v$ in terms of DNA chains per unit
%volume. The solution can also be imagined as a system of
%reflection images of one DNA in the walls of volume $v$. In either
%case we assume that the coil size $R$ is much smaller than the
%volume $v$ linear dimension, $R < v^{1/3}$. For the solution, this
%guarantees that coils do not overlap: $R^3 (1/v) < 1$.
Our argument for this case starts from noticing that there is a
cross-over spatial scale $\xi$, such that correlated sliding takes
place inside scale $\xi$, while regular 3D diffusion in water
occurs on a larger length scale, as it breaks correlations between
desorbtion and subsequent re-adsorbtion.  Thus, the time of one
tour of 3D diffusion is the mean first passage time into any one
of the $L/\lambda$ balls of the size $\xi$ (here $\lambda$ is the
contour length of DNA accommodated by one ball of the size $\xi$;
once again, we \emph{pretend} that we do not know $\xi$ and
$\lambda$, we will re-derive them in this single-protein
language).  The arrival time into one such ball is the
Smoluchowski time (discussed in the appendix
\ref{sec:Smoluchowski}) for the target of size $\xi$, it is about
$v/D_3\xi$; the arrival time into any one of the $L/\lambda$ balls
is $L/\lambda$ times smaller: $\sim v\left/ D_3 \xi (L/\lambda)
\right.$
%Our argument for this case starts from noticing that the
%tour of 3D diffusion begins once protein leaves a ball of size
%$\xi$ that houses the previous correlated sliding segment
%$\lambda$. Using DNA solution language, each DNA has $L/\lambda$
%such segments, and the 3D diffusion can be considered over as soon
%as protein arrives into any of the $L/\lambda$ balls of the size
%$\xi$ on any DNA. This means the time of 3D diffusion is just the
%Smoluchowski time for the ``target'' of the size $(L/\lambda)\xi$,
%this time is about $v \left/(L/\lambda)\xi D_3 \right.$.
%In other words, the protein has to diffuse in space over the
In order to present our equations for $\lambda$ and overall search
time $\tau$ in the form similar to Eqs.
(\ref{eq:ratio_of_times_in_globule}) and
(\ref{eq:full_time_for_globule}), we define distance $r_{\rm eff}$
such that $r_{\rm eff}^2 \sim D_3\left[ v\left/ D_3 \xi
(L/\lambda) \right.\right] = v\lambda/L\xi$ and then we obtain
%This finally yields equations for $\lambda$ (replacing
%(\ref{eq:ratio_of_times_in_globule}))
%
\be \frac{\lambda^2/D_1}{r_{\rm eff}^2/D_3} \sim \frac{y Lb^2}{v}
\ , \label{eq:ratio_of_times_in_coil} \ee
and
%for the overall search time (replacing (\ref{eq:full_time_for_globule}))
%
\be \tau \sim \frac{L}{\lambda} \left[ \frac{\lambda^2}{D_1} +
\frac{r_{\rm eff}^2}{D_3} \right] \ .
\label{eq:full_time_for_coil} \ee
Once again, remembering two regimes for the relation between
$\lambda$ and $\xi$, formula (\ref{eq:fractality}), and having
either first or second term dominate in the total time
(\ref{eq:full_time_for_coil}), we recover four regimes A, B, C,
and D.

Finally, the results for all saturation regimes E, F, G are
recovered by replacing the antenna length $\lambda$ with $L$ in
equation (\ref{eq:full_time_for_globule}) or
(\ref{eq:full_time_for_coil}), and replacing equality with
inequality in the conditions (\ref{eq:ratio_of_times_in_globule})
or (\ref{eq:ratio_of_times_in_coil}).

%Of course, the important step in our argument is an assumption
%that all $L/\lambda$ balls of the size $\xi$ each just add
%together in determining the effective target size.  In other
%words, they do not significantly screen each other.  This is
%equivalent to the assumption tacitly adopted earlier that the
%averaged flow lines in stationary diffusion approach go into the
%coiled antenna from the 3D space.  In fact, they do sometimes
%touch the other parts of the coil, but what we say is that the
%transport on the scale above $\xi$ is \emph{dominated} by the 3D
%diffusion, and, therefore, the screening between blobs of size
%$\xi$ is only marginal.  Of course, once there will a need to
%examine the corrections to scaling, this screening will have to be
%taken into account, along with the corrections to $D_1$ in the
%estimate of the correlated sliding time $\lambda^2/D_1$ due to the
%short excursions away from DNA.

\subsection{Comparison with earlier theoretical works}

Let us now compare our findings with various statements found in
the literature.  The most widely known result of the classical
work \cite{BWH} was the prediction, later confirmed experimentally
\cite{BWH2}, that the rate depends on $y$ (controlled by ionic
strength) in a characteristic way, exhibiting a maximum followed
by a plateau.   We have recovered this as a possible scenario for
some combinations of parameters (regimes), as shown in Fig.
\ref{fig:rate}. However, we found also a number of additional
features not noticed previously: first, the maximum is in many
cases asymmetric; second, the scaling of rate dependence on $y$
exhibits rich behavior, with the possibilities of crossing over
from $y^{1/2}$ to $y^{1/3}$ on the way to the maximum, or from
$y^{-2/3}$ to $y^{-1/2}$ on the way down; third, there is a
possibility of very strong deceleration at large adsorbtion
strength $y$ compared at the Smoluchowski rate.  All these
features have simple qualitative explanation: the rate grows
because increasing $y$ increases the antenna; the rate decays when
most of the proteins are fruitlessly adsorbed far from target (or,
in other language, every protein spends most of the time adsorbed
far away); the rate saturates and comes to the plateau because
antenna becomes as long as the DNA itself.  All of these features
are the direct consequence of the fractal properties of DNA, in
either coil or globule state.

The work Ref. \cite{Bruinsma} represents a review of a variety of
topics related to protein-DNA interactions, and the issue of
search rate is considered only briefly.  In the context, the work
Ref \cite{Bruinsma} provides an important insight, used above in
presenting the formula (\ref{eq:full_time_for_globule}), that
subsequent rounds of 1D search are performed on uncorrelated
pieces of DNA.  In other words, there exists a cross-over from
mostly correlated events, earlier combined into one ``correlated
sliding length $\lambda$'', to mostly uncorrelated ones.  In
accord with this insight, the search time is linear in DNA length
in the regime I.

In the paper Ref. \cite{Marko} antenna length was explicitly
identified with the sliding distance (that is, with the bare
sliding distance, earlier in this paper denoted as $\ell_{\rm
slide} \sim b \sqrt{yd}$), and then essentially formula
(\ref{eq:full_time_for_globule}) was used to determine the search
time.  This approach is perfectly valid as long as the antenna is
straight, $\lambda = \xi$, and $\lambda = \ell_{\rm slide}$, it
predicts the symmetric maximum of $J(y)$ dependence, but it should
not be used when DNA antenna is coiled.  For the globular DNA, the
approximation of straight antenna - implicit in the identification
of $\lambda$ with bare $\ell_{\rm slide}$ - is valid for the right
end of the regime A and for the regime D, while of course other
globular regimes require going beyond this approximation.

The main emphasis of the article Ref. \cite{Mirny} is on the role
of non-uniform sequence of DNA, which may lead to either
non-specific adsorbtion strength $y$, or 1D diffusion coefficient
$D_1$, or both to be ``noisy'' functions of coordinate on DNA.  In
their review of the uniform homopolymer case, Ref. \cite{Mirny}
employ formula equivalent to our Eqs.
(\ref{eq:full_time_for_globule}) or (\ref{eq:full_time_for_coil}),
but instead of the condition like
(\ref{eq:ratio_of_times_in_globule}) or
(\ref{eq:full_time_for_coil}) they minimize overall time with
respect to $\lambda$.  As we pointed out before, this approach is
valid within the cross-over corridor around the line $y \sim
v/Lb^2$.  In general, the idea to apply variational principle is
very interesting.  It can be generalized beyond the above
mentioned corridor if one minimizes the overall dissipation, which
is equivalent to energy minimization in terms of electrostatic
analogy, as we show in appendix \ref{sec:electrostatic_analogy}.
Of course, minimization of dissipation is equivalent to the
diffusion equation as long as diffusion is linear.  Alternatively,
one can also think, as emphasized in the work Ref. \cite{Marko},
that search mechanism was subject to optimization by biological
evolution.  To employ this idea, it is obviously necessary first
to understand the possible search scenario, or regimes, existing
in physics, and then, on the next stage, one could attempt
optimization with respect to the parameters, such as DNA packing
properties etc, which could be subject to selective pressure in
evolution.

BWH \cite{BWH} and some subsequent authors treated DNA solution in
terms of domains.  Although this term was never particularly
clearly defined, it could be understood as space regions more or
less occupied by separate DNA coils in solution.  With such
understanding, the terminology of domains can be used as long as
DNA coil fits into the volume $v$, or, in other words, better
suitable for an \emph{in vitro} experiment, DNA solution is
dilute, such that DNA coils do not overlap.  The terminology of
DNA domains becomes unsatisfactory at larger DNA concentrations.

Work Ref. \cite{FrenchGroup} considered the stochastic approach,
which means they did not look at the stationary diffusion, but
rather at the trajectory of a single protein.  As we pointed out
before, these approaches must be equivalent as long as one is only
interested in the average time of the arrival of the first of
proteins.  The important contribution of the work Ref.
\cite{FrenchGroup} was the elucidation of the crucial neglect of
the correlations between the desorbtion point of a protein and its
re-adsorbtion point.  It is because of this crucial and not always
justified approximation previous theories appear to have
overlooked the mechanism of correlated re-adsorbtion, which is
entirely due to the DNA being a polymer and a fractal coil.
Correlated re-adsorbtion was anticipated in the experimental works
\cite{Halford}.

\subsection{Experimental situation}

Most of the experiments in the field (see review \cite{Halford}
and references therein) involve various ingenious arrangements of
two or more target sites on the linear or ring DNA and observation
of the resulting enzyme processivity.  In the light of our theory,
it would be interesting to revive the earlier BWH-style
experiments and to look carefully at the theoretically predicted
multiple features of $J(y)$ curves, such as asymmetric maximum,
various scaling regions, the possible deceleration, etc.

The seeming difficulty is that all our ``interesting'' regimes
start when $y > p^2/b^2d$, when antenna is longer than DNA
persistence length.  Since persistence length of dsDNA, $p$, is
fairly large, about 150 base pairs under usual ionic conditions
(say, $\left[ {\rm Na} \right] = 0.2 \ {\rm M}$), and assuming $b$
is about the diameter of the double helix, we get $p/b \approx 25$
for the dsDNA.  Unless $d$ is large, this seems to require fairly
large non-specific adsorbtion energies, about $6k_BT$ to $10k_BT$,
which is a lot but not impossible.  In any case, we would like to
emphasize that the maximum $J(y)$ \emph{has} been observed
\cite{BWH2}, which, according to our theory, could have happened
only at $y > p^2/b^2d$, thus assuring that this range is within
reach.

One of the most critical and poorly known parameters of our theory
is $d= D_1/D_3$. Of course, $D_3$, diffusion coefficient of the
protein in water, is known pretty well, and can be simply
estimated based on its size using Stokes-Einstein relation.  The
difficult part is about $D_1$, which involves friction of the
protein against DNA in the solvent.  It is clear that slow
diffusion along DNA would make the entire mechanism of 1D sliding
less efficient, and indeed decreasing $d$ systematically reduces
the rate that we obtain in almost all regimes.  There are only two
exceptions to this: one is trivial, it is pure Smoluchowski
process not involving any sliding and realized only when there is
no non-specific adsorbtion on DNA ($y \leq 1$); another exception
is in the regimes E and F - regimes when entire DNA, rod-like or
coil-like, serves as an antenna, which means 3D transport to the
DNA is the slowest part, the bottleneck of the whole process, so
that reducing $d$ does not do any damage - except, of course,
pushing away the corresponding regime boundaries.

Experimental data on the 1D diffusion of proteins along DNA are
scarce and not completely clear \cite{Eber}.  %Very recently, when

An interesting spin on the whole issue of 1D transport is added by
the proteins, such as, e.g., helicase, which, provided with proper
energy supply, can move actively.  For us, in the context of our
present theory, active movement is likely to correspond to great
increase of $D_1$, or $d$, for either actively moving proteins
themselves, or for passively diffusing proteins which might
receive push or pull from active ones.  At the first glance, this
sounds like a paradoxical statement, because active motion is not
diffusion in the sense that displacement is linear in time.
However, this is only true up to a certain time and length scales.
At larger scale, we can reasonably assume that it would be
diffusion again, albeit with a vastly increased diffusion
coefficient.  Indeed, first, there is always a probability of
thermally activated detachment from DNA, and, second, given that
two strands in DNA are antiparallel, the re-adsorbtion is likely
to lead to random choice of direction of further sliding.  These
two ingredients surely correspond to diffusion, in the sense that
displacement goes like $t^{1/2}$.  Of course, this entire issue of
active transport requires further investigation, which naturally
brings us to the conclusion of this paper.

\section{Conclusion}

Many questions remain open.  The role of concurrent protein
species, the role of non-uniform DNA sequence, the role of DNA
motion \cite{Berg_moving_DNA}, the probability of unusually long
search times, the search on a single stranded DNA or RNA, the role
of superhelical structures, the dependence of rate (or search
time) on the specific positions of one or more targets on DNA, the
related issue of enzyme processivity, the role of excluded volume
for very long DNA and corresponding loop-erasing walks
\cite{loop_erasing_walk} - all of these questions invite
theoretical work.

To conclude, we have analyzed all scaling regimes of the
diffusion-controlled search by proteins of the specific target
site located on DNA.  We found many regimes.  The major idea can
be formulated in terms of the cross-over between 1D sliding along
DNA up to a certain length scale and 3D diffusion in surrounding
space on the larger length scale.  Overall, qualitatively, this
idea seems to be in agreement with the intuition expressed in
experimental papers.  In addition, we have made several
theoretical predictions which are verifiable and (even more
importantly) falsifiable by the experiments.  We are looking
forward to such experiments.

\begin{acknowledgments}  We gratefully acknowledge very useful discussion with
J.-L. Sikorav.  The work of AYG was supported in part by the MRSEC
Program of the National Science Foundation under Award Number
DMR-0212302. \end{acknowledgments}

\appendix

\section{Simple scaling derivation of the Smoluchowski
rate and the Smoluchowski time}\label{sec:Smoluchowski}

Classical Smoluchowski theory \cite{Smoluchowski} treats the
diffusion-controlled process of irreversible absorbtion of
diffusing particles by an immobile sphere of a given radius, call
it $b$.  As in our proteins problem, Smoluchowski theory can be
formulated either in terms of stationary rate $J_s$, assuming
concentration $c$ is fixed, or in terms of mean first passage time
$\tau_s$ for a single protein.

Let us imagine that a protein diffuses within a volume $v$, and
its diffusion coefficient is $D_3$.  Let us further define the
time interval $t_b$ such that over time $t_b$ protein moves the
distance of order $b$:  $D_3 t_b \sim b^2$.  Then, over a longer
time $t$ protein visits $t/t_b$ spots of the size $b$ each, and,
given that $b^3 \ll v$, the probability that none of this spots is
the target, or the probability to keep missing target for the time
$t$ obeys Poisson distribution and decays exponentially with $t$:
$\left( 1 - b^3/v \right)^{t/t_b} \simeq \exp \left[ - t b^3 / (v
t_b) \right]$. The mean first passage time is read out of this
formula, it is $\tau_s \sim v /(D_3 b )$.

The corresponding stationary rate is obtained by inverting this
time, assuming overall concentration of proteins $c =1/v$.  Thus,
$J_s \sim D_3 c b$.

Of course, more accurate derivation, available in a number of
textbooks (and easily formulated in terms of electrostatic
analogy, see section \ref{sec:electrostatic_analogy}), is
necessary to complement the result with the correct prefactor of
$4 \pi$.

\section{Electrostatic analogy}\label{sec:electrostatic_analogy}

Here, we re-derive the results of the section
\ref{sec:simple_case} using the fact that stationary diffusion
equation is the same as Laplace equation in electrostatics.
Specifically, the problem of diffusion into the target of the size
$b$ is equivalent to the problem of finding the electric field
around a charge of the size $b$.  The key relatively non-trivial
point of this analogy is to realize that the potential well for
diffusing particles is equivalent in electrostatic language to the
region in space with very high dielectric constant.  In our case
the potential well is located all around DNA, and the target is
also somewhere on the DNA.  Therefore, it is equivalent to the
electrostatic problem in which we have a channel, of the diameter
about $b$, filled with high dielectric constant material, for
instance - water, and surrounded by a low dielectric constant
material.  Specifically, it is easy to check that $y$ of the
diffusion problem is exactly equivalent to $\epsilon_w/\epsilon_m$
- the ratio of dielectric constants of water and surrounding
medium:  $y=\epsilon_w/\epsilon_m \gg 1$.

Thus, we have to address the problem of a charge $Q$ located
inside the water filled channel in, let say, a thick lipid
membrane. For the straight channel, this is a well known problem
in membrane biophysics. It was first studied by Parsegian
\cite{Parsegian}, and the recent most detailed exposition is given
in the article \cite{ZhangKamenevLarkinShklovskii}.  Here, we give
only simple scaling consideration.

Since $\epsilon_w/\epsilon_m \gg 1$, field lines prefer to remain
inside the channel for as long as possible.  This gives the
picture of electric field equivalent to the Fig.
\ref{fig:antenna}, a or b.  In other words, we should say that
there is some length scale $\lambda$ along the channel, and within
this scale electric field lines are predominantly confined in the
channel.  At the same time, outside of the sphere of radius $\xi$,
electric field is close to that of a spherical charge in
unrestricted space.  Thus, electric field energy can be
approximated as the sum of two parts, one due to the uniform field
in the volume about $b^2\lambda$ in the channel, and the other
around the $\xi$-sphere in the medium.  Since $E$-field in the
channel is about $Q/b^2 \epsilon_w$ while $D$-field is $Q/b^2$,
the part of energy due to the field inside the channel is about
$\left( Q/b^2 \epsilon_w \right) \times \left(Q/b^2 \right) \times
\left(b^2 \lambda \right) = Q^2 \lambda / b^2 \epsilon_w$.  At the
same time, energy of the field in the outer zone is about $Q^2/\xi
\epsilon_m$.  Thus, total electrostatic energy (self-energy of the
charge $Q$) is
\be E \sim \frac{Q^2 \lambda}{b^2 \epsilon_w} + \frac{Q^2}{\xi
\epsilon_m} \ . \label{eq:electrostatic energy} \ee

To begin with, let us assume that the channel is straight.  Then,
$\lambda = \xi$, and minimization of the energy
(\ref{eq:electrostatic energy}) gives $\lambda \sim b
\sqrt{\epsilon_w/\epsilon_m} \gg b$.  This formula can be found in
the book ref. \cite{Finkelstein_Ptitsyn}.  Given that
$y=\epsilon_w/\epsilon_m$, this formula is equivalent to our
result for the antenna length in the straight antenna regime A
(assuming $d=1$).

Consider now coiled channel; such problem was never considered in
electrostatic context, but one can imagine, for instance, a
flexible fiber of high dielectric constant material surrounded by
air. Formula (\ref{eq:electrostatic energy}) still applies, but
$\xi \sim \sqrt{\lambda p}$. Minimization then yields $\lambda
\sim b^{4/3} p^{-1/3} \left( \epsilon_w / \epsilon_m \right)^{2/3}
= b^{4/3} p^{-1/3} y^{2/3}$, which is our result for the antenna
length in the regime B.

To conclude, we note that minimization of energy in the
electrostatic language is translated to minimization of
dissipation in the diffusion language.

\end{document}